\documentstyle[aps,prb,amsmath,amssymb,epsfig]{revtex}
\begin{document}
\title{Conductivity of disordered electrons: Mean-field approximation
containing vertex corrections}
\author{V. Jani\v{s}$^1$ and D. Vollhardt$^2$}
\address{$^1$ Institute of Physics, Academy of Sciences of the Czech Republic,\\
Na Slovance 2, CZ-18221 Praha 8, Czech Republic\\
$^2$
Theoretische Physik III, Elektronische Korrelationen und\\
Magnetismus, Institut f\"ur Physik, Universit\"at Augsburg,\\
D-86135 Augsburg, Germany}
\date{\today}
\maketitle

\begin{abstract}
The electrical {\em dc}-conductivity of disordered, non-interacting
electrons is calculated in the asymptotic limit of high lattice dimensions $%
d\rightarrow \infty $. To go beyond the lowest-order contribution in the
expansion parameter $1/d$ of the single bubble diagram, vertex corrections
are calculated from an asymptotic expression
for the two-particle vertex. A mean-field approximation for the $%
dc$-conductivity containing the leading high-dimensional vertex corrections
is proposed which is free of spurious non-analyticities, i.e. the
conductivity is non-negative and shows no unphysical behavior in $d\geq 3$.
\end{abstract}





Electronic problems with interactions or disorder can almost never
be solved exactly, except for special limits. The resolvent
functions of an exact solution fulfill conservation laws and have
the correct analytic (Herglotz) properties, i.e. do not contain
spurious poles. This is generally not true for approximate, e.g.
perturbative, solutions. Only global, self-consistent
approximations have a chance to be free of unphysical behavior and
to yield the desired analyticity of a solution for all input
parameters.

For non-interacting tight-binding electrons in a random potential the first
self-consistent solution with the correct analytic properties was the
''coherent potential approximation'' (CPA). \cite{Velicky68,Elliot74} The
Herglotz analyticity of the CPA-equations was proved explicitly by
M\"{u}ller-Hartmann. \cite{Mueller-Hartmann73} Only later the CPA was found
to be the {\em exact} solution of the Anderson disorder model in two
particular limits. First, the CPA with a semicircular disorder distribution
was shown to correspond to the exact solution of an $n$-orbital model in the
limit $n=\infty. $\cite{Wegner79} Then, after the limit of high spatial
dimensions $d$ for fermionic lattice models had been introduced, \cite
{Metzner89} it was found that the CPA represents the exact solution of the
Anderson disorder model in $d=\infty $ for arbitrary disorder distributions.
\cite{Vlaming92,Janis92} Since then the limit $d\rightarrow \infty $ has
served as a useful tool for deriving self-consistent, fully dynamical
approximation schemes for interacting lattice electron systems, \cite
{Georges96} referred to as Dynamical Mean-Field Theory (DMFT).

By calculating a physical quantity in $d=\infty $ one obtains a
particular mean-field value. The situation becomes subtle if the value
obtained in this limit is zero. This is, for example, the case for {\em %
non-local} quantities such as the off-diagonal propagator $G_{ij}$, $i\neq j$%
. They depend on the distance between two or more different lattice sites
(i.e., their Fourier transform is wave-vector dependent) and are thus
necessarily proportional to some power of $1/d$, e.g., $G_{ij}\sim O(1/\sqrt{%
d})$, for nearest-neighbor sites $i,j$. However, that does not imply that
these quantities can be neglected in $d=\infty $. They may contribute, since
they appear in lattice sums where the summation over the $d\rightarrow
\infty $ many sites compensates their $1/d$-smallness. To include these
quantities properly, one has to calculate their asymptotic behavior in the
limit $d\rightarrow \infty $, thus going beyond the strict $d=\infty $
limit. The necessity to go beyond $d=\infty $ is also evident from the fact
that higher-order non-local Green functions are related to lower-order local
Green functions\ by functional derivatives via generalized Ward identities,
reflecting conservation laws. It was recently shown by one of us \cite
{Janis99a} and by Hettler et al. \cite{Hettler99} that, when only the
leading asymptotic contributions to one- and two-particle Green functions
are taken into account, the Ward identities are not fulfilled in $d=\infty $%
. One has to go beyond the leading order at the lower particle level, or
introduce anomalous functions, to restore conservation laws. \cite{Janis99a}
This shows that the definition of two- and higher-order Green functions is
ambiguous in the limit $d=\infty $.

A particularly important example of a quantity whose mean-field value in $%
d=\infty $ vanishes, is the electrical conductivity $\sigma $. It is defined
from a Kubo formula with the current-current correlation function. In the
limit $d\rightarrow \infty $ the optical $dc$-conductivity is given by a
single bubble diagram, with $\sigma \thicksim O(1/d)$. \cite{Khurana90} In a
formal $1/d$-expansion this result would be the first non-vanishing
contribution to $\sigma $. It is useful to consider this {\em non-vanishing
asymptotic result} as a "mean-field value" of the conductivity \cite{Note1}.
Likewise one may define a mean-field value of any physical quantity through
its leading non-vanishing asymptotic result in the limit $d\rightarrow
\infty $.

A mean-field result for the $dc$-conductivity $\sigma \thicksim O(1/d)$
defined in this way does not contain vertex corrections. Hence it
does not include the physics of back-scatterings. However, in random systems
vertex corrections are known to be extremely important since they are
responsible for Anderson localization at zero temperature in sufficiently
low dimensions ($d=1,2$) or for sufficiently strong disorder in $3\leq
d<\infty $. At least from a diagrammatic point of view it is not yet fully
understood how the conductivity $\sigma $ approaches zero at the
localization transition. \cite{Lee85,Vollhardt92} Clearly one has
to go beyond the mean-field single-site diagrams to incorporate localization
effects.

Most recently Jarrell and Krishnamurthy \cite{Jarrell+Krish00} introduced
systematic non-local corrections to the CPA on the one-particle level using
the Dynamical Cluster Approximation to obtain results compatible with
Herglotz analyticity, i.e., a non-negative density of states. Here we choose
another route to go beyond the mean-field limit and calculate non-local
corrections to the CPA two-particle {\it irreducible} vertex function. The
aim of our paper is to employ the limit of high lattice dimensions to
improve upon the mean-field conductivity $\sigma \thicksim O(1/d)$ by
including vertex corrections. We follow the proposal of ref.~\cite{Janis99a}
where the high-dimensional asymptotics of the full vertex function with
leading vertex corrections to the electrical conductivity was derived. Our
main result in this paper is a mean-field expression for the electrical
conductivity which includes leading asymptotic vertex corrections in $%
d\rightarrow \infty $ while remaining non-negative in $d\geq 3$.

The vertex function from ref.~\cite{Janis99a} contains the leading
$1/d$-asymptotics of all two-particle quantities. Employing the
Kubo formula for the electrical conductivity with the two-particle vertex we
may derive the leading asymptotics for the conductivity and its vertex
corrections. However, approximations of the full vertex function in the Kubo
formula can, in principle, lead to unphysical results. Indeed, the vertex
corrections to the single-bubble term may have a negative sign and hence
positivity of the conductivity cannot be warranted. Clearly, a meaningful
and consistent approximation for the conductivity must never become
negative. To obtain such an approximation we represent the full vertex
function by means of an irreducible vertex and a Bethe-Salpeter equation in
the electron-hole channel. To produce the leading contribution to the vertex
corrections in $d\rightarrow \infty $, the irreducible function must be
evaluated with its leading {\em non-local} contribution. By simplifying the
Bethe-Salpeter equation in high spatial dimensions we then obtain a closed,
mean-field expression for the conductivity with vertex corrections.

\bigskip In the following we consider the Anderson disorder Hamiltonian
\begin{eqnarray}  \label{eq:AD_hamiltonian}
H =-\frac{t^{\ast }}{\sqrt{Z}}\sum_{<ij>}c_{i}^{\dagger} c_{j} +\sum_{i}
V_{i}c_{i}^{\dagger }c_{i}\ ,
\end{eqnarray}
to describe the effects of randomness. Here $t^{\ast }$ is the hopping
matrix element between nearest neighbors, scaled in such a way as to produce
a meaningful limit $d\rightarrow \infty $, with $Z$ as the coordination
number of the lattice. \cite{Metzner89} The local, static potential $V_{i}$
is a random variable with site-independent distribution function. The
conductivity of a quenched random system without interparticle interactions
is described by averaged one- and two-particle Green functions (resolvents) $%
G_{ij}(z)=\langle \left[ z\widehat{1}-\widehat{t}-\widehat{V}\right]_{ij}
\rangle _{av}$ and $G^{(2)}_{ij,kl}(z_{1},z_{2})=\left\langle \left[ z_{1}%
\widehat{1}-\widehat{t}-\widehat{V}\right] _{ij}^{-1}\left[ z_{2}\widehat{1}-%
\widehat{t}-\widehat{V}\right] _{kl}^{-1}\right\rangle _{av}$, respectively.
It is our first goal to determine these functions in the asymptotic limit $%
d\rightarrow \infty $.

It is straightforward to derive the $d\rightarrow \infty $ limit
of the self-energy which carries the information about how the
randomness influences the motion of a single electron. The
self-energy becomes local and can be obtained from the single-site
equation\begin{eqnarray}\label{eq:CPA_equation}
 \left\langle\frac
    1{1+\left(\Sigma(z)-V_i\right)G(z)}\right\rangle_{av}=1\ .
\end{eqnarray}
 This is precisely the well-known CPA
equation\cite{Velicky68} for the self-energy. Here the local
(diagonal) one-particle propagator is denoted by
$G(z)=N^{-1}\sum_{{\bf k}}G({\bf k},z)=\int d\rho (\epsilon
)[z-\Sigma (z)-\epsilon ]^{-1}$ where $\rho $ is the density of
states. It is less evident how to derive expressions for averaged
two-particle functions which are consistent with the local
self-energy, since we have to work explicitly with non-local
quantities. At the two-particle level one has to keep two separate
lattice points to derive the leading asymptotics for large $d$.
\cite {Janis99a} It is more convenient and practical to work with
an averaged cumulant, or better with a vertex $\Gamma $ defined in
momentum space as
\begin{eqnarray}
\Gamma ({\bf k}_{1},z_{1},{\bf k}_{2},z_{2};{\bf q})&=&G^{-1}({\bf k}%
_{1},z_{1})G^{-1}({\bf k}_{2},z_{2})  \nonumber  \label{eq:2P_vertex}
\left[ G^{(2)}({\bf k}_{1},z_{1},{\bf k}_{2},z_{2};%
{\bf q})-\delta ({\bf q})G({\bf k}_{1},z_{1})G({\bf k}_{2},z_{2})\right]
\nonumber \\ &&\hspace*{15mm}\times 
G^{-1}({\bf k}_{1}+{\bf q},z_{1})G^{-1}({\bf k}_{2}+%
{\bf q},z_{2})\ .
\end{eqnarray}

It was shown in ref.~\cite{Janis99a} that the asymptotic $d\rightarrow
\infty $ solution for the vertex $\Gamma $ can be represented as a
sum of three contributions, solutions of Bethe-Salpeter equations
in three inequivalent two-particle irreducibility channels. The
integral kernels of these equations are always the local
two-particle irreducible vertex (being the same in all channels)
accompanied by a non-local two-particle bubble. The local
two-particle irreducible vertex in $d=\infty $ reads
\begin{eqnarray}\label{eq:2IP_vertex} &&\Lambda
(z_{1},z_{2})=\frac{\delta \Sigma (z_{1})}{\delta
G(z_{2})}=\frac{1}{G(z_{1})G(z_{2})}\left[ 1
-\left\langle \frac{1}{1+\left( \Sigma (z)-V_{i}\right) G(z_{1})}%
\frac{1}{1+\left( \Sigma (z)-V_{i}\right) G(z_{2})}\right\rangle _{av}^{-1}%
\right]
\end{eqnarray}
and the
two-particle bubbles containing the entire momentum dependence are
given by $\chi ^{\pm }({\bf q};z_{1},z_{2})=N^{-1}\sum_{{\bf k}}G({\bf k}%
,z_{1})G({\bf \ q}\pm {\bf k},z_{2})$. In the limit $d\rightarrow \infty $
the vertex functions from the three two-particle channels (electron-hole,
electron-electron, and vertical) take the form\cite{Janis00}
\begin{subequations}  \label{eq:infty_channels}
\begin{eqnarray}
  \label{eq:infty_eh-channel}
  \Gamma^{eh}({\bf q};z_1,z_2)&=& \frac{\Lambda(z_1,z_2)}
  {1-\Lambda(z_1,z_2)\chi^+({\bf q};z_1,z_2) }\ ,  \\ \label{eq:infty_ee-channel}
   \Gamma^{ee}({\bf q};z_1,z_2)&=& \frac{\Lambda(z_1,z_2)}
  {1-\Lambda(z_1,z_2) \chi^-({\bf q};z_1,z_2)}\ ,  \\
  \label{eq:infty_v-channel}
  \Gamma^{v}({\bf q};z_1,z_2)&=&\gamma(z_1,z_2) \prod_{i=1}^2
  \frac{1-\Lambda(z_i,z_i)G(z_i)G(z_i)}
  {\left[1-\Lambda(z_i,z_i)\chi^+({\bf q};z_i,z_i) \right]} \ .
\end{eqnarray}
\end{subequations}
The local part of the vertex functions in (\ref{eq:infty_channels}) is
always the same, i.e. is given by $\gamma (z_{1},z_{2})=\Lambda
(z_{1},z_{2})/\left[ 1-\Lambda (z_{1},z_{2})G(z_{1})G(z_{2})\right] $. The
full vertex is a sum of the above three contributions, where the
transferred momentum ${\bf q}$ has a different meaning in each channel.
This is due to the fact that the irreducibility channels are topologically
inequivalent and differ in the momentum that is conserved during multiple
scatterings. If the incoming particle and hole carry momenta ${\bf k}_{1}$
and ${\bf k}_{2}$ then the conserved momentum is ${\bf k}_{2}-{\bf k}_{1}$,
${\bf \ k}_{1}+{\bf k}_{2}+{\bf q}$, and ${\bf q}$ for the electron-hole,
electron-electron, and vertical channels, respectively. \cite{Janis99b} The
momentum ${\bf q}$ is the momentum transferred during the scattering on
impurities, i.e., the outgoing particle and hole carry momenta ${\bf \ k}
_{1}+{\bf q}$ and ${\bf k}_{2}+{\bf q}$, respectively. In order to avoid
multiple summation on the same site we must subtract the local vertex from
the sum of the channel-dependent vertex functions twice. We then obtain an
explicit representation for the two-particle vertex in high dimensions in
the notation of ref.~\cite{Janis99a}
\begin{eqnarray}
  \label{eq:d_infty_vertex}
&&\Gamma ({\bf k}_{1},z_{1},{\bf k}_{2},z_{2};{\bf q})=\Gamma
^{eh}({\bf k}_{2}-{\bf k}_{1};z_{1},z_{2})
+\Gamma ^{ee}({\bf k}_{1}+{\bf k}_{2}+{\bf q}
;z_{1},z_{2})+\Gamma ^{v}({\bf q};z_{1},z_{2})-2\gamma
(z_{1},z_{2})\ .
\end{eqnarray}
We note that the CPA vertex function derived in ref.~\cite{Velicky69} %
is given by only the first term in the above equation, i.e.  $\Gamma
^{eh}$. Hence it does not contain the transferred momentum $%
{\bf q}$ needed to incorporate vertex corrections to the
conductivity.

The density of the static ($dc$) electrical conductivity at zero temperature
is defined by a Kubo formula with the full vertex as ($\hbar=1$)
\begin{eqnarray}  \label{eq:T=0_conductivity}
&&\mbox{Re}\ \sigma_{\alpha\beta}=\frac{e^2}{4\pi}\frac 1{N^2}\sum_{{\bf k},%
{\bf k}^{\prime}} v_\alpha({\bf k}) v_\beta({\bf k}^{\prime})\sum_{\sigma%
\tau} (-\sigma\tau) G_\sigma({\bf k}) G_\tau({\bf k})  
\left[\delta({\bf k}-{\bf k}^{\prime})+\
\Gamma_{\sigma\tau}({\bf k},{\bf k}; {\bf k}^{\prime}-{\bf k}) G_\sigma({\bf %
k}^{\prime}) G_\tau({\bf k}^{\prime})\right]
\end{eqnarray}
where $\sigma ,\tau =\pm 1$, $\Gamma _{\sigma \tau }({\bf k},{\bf \ k}%
^{\prime };{\bf q})=\Gamma ({\bf k},E_{F}+i\sigma 0^{+},{\bf k}^{\prime
},E_{F}+i\tau 0^{+};{\bf \ q})$, $G_{\sigma }({\bf k})=G({\bf k}%
,E_{F}+i\sigma 0^{+})$, $v_{\alpha }({\bf k})=m^{-1}\partial \epsilon ({\bf k%
})/\partial k_{\alpha }$, $\epsilon ({\bf k})$ is the dispersion relation
and $m$ the mass of the electron.

Eq.~(\ref{eq:T=0_conductivity}) with the vertex functions (\ref
{eq:infty_channels}) contains nontrivial corrections to the one-electron
conductivity (the single-bubble diagram). However, (\ref{eq:T=0_conductivity}%
) is not appropriate for approximate evaluations in finite
dimensions. The vertex corrections are merely {\em added} to the
one-particle conductivity, such that negative contributions may
reverse the overall sign, thus leading to unphysical behavior. To
avoid such a situation we represent the conductivity in a
different way. We use a Bethe-Salpeter equation in the
electron-hole channel expressing the full vertex $\Gamma $ via an
irreducible one, $\Lambda ^{eh}$. The irreducible vertex $\Lambda
^{eh}$ together with $\chi ^{+}$ determine the
integral kernel of the Bethe-Salpeter equation explicitly and
define a matrix multiplication scheme in momentum space.
\cite{Janis99b} The integral kernel and the multiplication rule
for the electron-hole channel are given by
\begin{subequations}
\begin{eqnarray} \label{eq:eh_kernel}
  \left[\Lambda^{eh}_{\sigma\tau}G_\sigma G_\tau\right]({\bf k},{\bf
    k}';{\bf q}) &=& \Lambda^{eh}_{\sigma\tau}({\bf k},{\bf k}';{\bf q})
  G_\sigma({\bf k}+{\bf q}) G_\tau({\bf k}'+{\bf q})\ ,\\ \label{eq:eh_multiplication} 
 \left[X\bullet Y\right]({\bf k},{\bf k}';{\bf q})&=&  \frac 1{N}\sum_{{\bf q}'}
  X({\bf k},{\bf k}';{\bf q}') Y({\bf k}+{\bf q}',{\bf k}'+{\bf q}';{\bf
    q}-{\bf q}')\ .
\end{eqnarray}
\end{subequations}
A solution of the
Bethe-Salpeter equation for the two-particle vertex
$\Lambda ^{eh}$ can formally be written as $\Gamma =\left\{ 1-%
\left[ \Lambda ^{eh}GG\right] \bullet \right\} ^{-1}\Lambda ^{eh}$
where the bullet indicates that, upon expansion of $\left\{
...\right\} ^{-1}$, the two-particle functions $\left[ \Lambda
^{eh}GG\right] $ are multiplied according to (\ref
{eq:eh_multiplication}). Inserting this solution into (\ref
{eq:T=0_conductivity}) we obtain a new, equivalent representation
for the conductivity \begin{eqnarray}
  \label{eq:cond_eh_repr}
  \mbox{Re}\ \sigma_{\alpha\beta}&=&\frac{e^2}{4\pi}\frac 1{N^2}\sum_{{\bf
      k},{\bf k}'} v_\alpha({\bf k})v_\beta({\bf k}')\sum_{\sigma\tau}
  (-\sigma\tau)  G_\sigma({\bf k}) G_\tau({\bf k})\left\{1
    -\left[\Lambda^{eh}_{\sigma\tau}G_\sigma G_\tau\right]\bullet\right\}^{-1} ({\bf
  k},{\bf k};{\bf k}'-{\bf k}) \ .
\end{eqnarray}
For not too strong disorder the norm of the
operator $\left\| \Lambda _{\sigma \tau }^{eh}G_{\sigma }G_{\tau
}\right\| \lesssim 1$. Hence the conductivity remains
non-negative.

Note that only the non-local part with odd parity with respect to
reflections in ${\bf k}$ and ${\bf k}^{\prime }$ of the vertex $\Lambda
^{eh}({\bf k},{\bf k};{\bf k}^{\prime }-{\bf k})$ contributes to the
conductivity. We obtain its leading asymptotic term if we use the
representation $\Gamma =\left\{ 1-\left[ \Lambda ^{eh}GG\right] \bullet
\right\} ^{-1}\Lambda ^{eh}$ and solve it for $\Lambda ^{eh}$, i.e., $%
\Lambda ^{eh}=\Gamma \left\{ \bullet \left[ GG\Gamma \right]
+1\right\} ^{-1} $. Using the vertex $\Gamma $ from
(\ref{eq:d_infty_vertex}) one finds in the order $O(1/d)$
\begin{eqnarray}
  \label{eq:decoupling}
\Lambda ^{eh}({\bf k}_1,z_1,{\bf k}_2,z_2;{\bf
  q})&=&\Lambda(z_1,z_2)
+\left(1-\Lambda(z_1,z_2)G(z_1)G(z_2) \right)^2\left[\Gamma ({\bf
    k}_{1},z_{1},{\bf k}_{2},z_{2};{\bf q})- \Gamma^{eh} ({\bf k}_{2}- {\bf
    k}_{1};z_{1},z_{2})\right] .
\end{eqnarray}
The irreducible vertex (\ref{eq:decoupling}) together with the
multiplication scheme (\ref{eq:eh_multiplication}) used in (\ref
{eq:cond_eh_repr}) leads to an integral-equation representation of
the conductivity. In the limit $d\rightarrow \infty $ the momentum
convolutions decouple. This fact helps us further simplify the
expression for the conductivity. To derive the leading
asymptotic contribution from the nonlocal part of $\Lambda ^{eh}$
to the conductivity we have to
calculate the momentum convolutions on the level of order $%
O(1/d) $ so that the velocities appear in squares and the momentum integrals
do not vanish.

In the following we resort to a hypercubic lattice where only the diagonal
(longitudinal) conductivity remains. Keeping only the leading-order terms in
the expansion of the denominator in (\ref{eq:cond_eh_repr}) we end up with a
mean-field-like expression for the {\em dc}-conductivity
\begin{subequations}\label{eq:mf_cond}
\begin{eqnarray}
  \label{eq:cond_positive}
  \mbox{Re}\ \sigma_{\alpha\alpha}=\frac{e^2}{4 \pi}
  \sum_{\sigma\tau}(-\sigma\tau) \frac{\langle v_\alpha^2 G_\sigma
    G_\tau\rangle }{1-\langle v_\alpha^2G_\sigma G_\tau\rangle
    \langle\Lambda^{\prime\alpha}_{\sigma\tau} \rangle }
\end{eqnarray}
where $\langle v_\alpha^2 G_\sigma G_\tau\rangle = N^{-1}\sum_{\bf
  k}v_\alpha({\bf k})^2  G_\sigma({\bf k})G_\tau({\bf k})$ and
\begin{eqnarray}
  \label{eq:Lambda_prime}
  \langle\Lambda^{\prime\alpha}_{\sigma\tau}\rangle&= &\frac
  1{N^{2}}\sum_{{\bf k},{\bf k}'}
  \frac{\delta^2} {\delta v_\alpha({\bf  k})\delta v_\alpha({\bf k}')}
  \Lambda^{eh}_{\sigma \tau}({\bf k},{\bf k};{\bf k}'-{\bf k}) \ .
\end{eqnarray}\end{subequations}
In the asymptotic limit $d\rightarrow \infty $ the irreducible vertex $%
\Lambda ^{eh}$ is determined from (\ref{eq:infty_channels}) and (\ref
{eq:decoupling}) with a simplified momentum dependence of $\Lambda ^{eh}$
via the quantity $X({\bf k})=\frac{1}{d} \sum\limits_{\nu =1}^d \cos k_{\nu
}$. \cite{Mueller-Hartmann89} We note that in the asymptotic limit $%
d\rightarrow \infty $ one has $\langle \Lambda ^{\prime }{}_{\sigma \tau
}^{\alpha }\rangle \sim O(1)$ and $\langle v_{\alpha }^{2}G_{\sigma }G_{\tau
}\rangle \sim O(1/d)$. For $\langle \Lambda ^{\prime }{}_{\sigma \tau
}^{\alpha }\rangle=0$ (as, for example, in the case of a ${\bf k}$%
-independent $\Lambda _{\sigma \tau }^{eh} $) the resulting
expression for the conductivity reduces to the CPA-result
\begin{eqnarray}
  \label{eq:sigma_CPA}
\mbox{Re}\ \sigma _{\alpha \alpha }^{\text{CPA}}=\frac{e^{2}}{4\pi
} \mathrel{\mathop{\sum }\limits_{\sigma \tau }} (-\sigma \tau
)\langle v_{\alpha }^{2}G_{\sigma }G_{\tau }\rangle.
\end{eqnarray}
This is
precisely the mean-field conductivity defined from the
$d\rightarrow \infty $ limit, with $\mbox{Re}\ \sigma _{\alpha
\alpha }^{\text{CPA}}\sim O(1/d)$ due to $v_{\alpha }^{2}\sim
O(1/d)$. The denominator in (\ref {eq:cond_positive}) then
contains the leading asymptotic contribution from the vertex
corrections to the conductivity.

We note that the self-energy $\Sigma $ and the vertex function $\Lambda
^{eh} $ are connected via a Ward identity. Velick\'{y} \cite{Velicky69}
showed that the CPA self-energy $\Sigma $, (\ref{eq:CPA_equation}), and the
CPA-vertex $\Lambda $, (\ref{eq:2IP_vertex}), fulfill the Ward-identity
exactly. In our case, where we use the local CPA self-energy and the
non-local part of the vertex function $\Lambda ^{eh}$, (\ref{eq:decoupling}%
), or rather $\langle \Lambda ^{\prime }{}_{\sigma \tau
}^{\alpha }\rangle $ from (\ref{eq:Lambda_prime}), the Ward identity is
fulfilled only asymptotically in the leading order of $1/d$ for
both quantities. This is fully consistent with the spirit of the
simplification we made in deriving the mean-field expression for the
conductivity with vertex corrections (\ref{eq:mf_cond}).

Generally, in (\ref{eq:mf_cond})  we have to perform integrals in
momentum space that cannot be reduced to integrals over the
density-of-states as it would be typical for mean-field theories.
A reduction to an expression with integrals over the
density-of-states is possible only if we resort to the leading
contribution to $\langle \Lambda ^{\prime }{}_{\sigma \tau
}^{\alpha }\rangle $ in $1/d$. This further simplification yields
\begin{eqnarray}
  \label{eq:cond_dos}
\mbox{Re}\ \sigma_{\alpha\alpha} &=& \left(
\frac{e^{2}t^{*2}}{8\pi
    d}\right) 
\sum_{\sigma\tau} \frac{(-\sigma\tau)  \langle G_\sigma
G_\tau\rangle} {1+\frac{t^{*2}}{2d}   \langle G_\sigma
G_\tau\rangle
  \Lambda_{\sigma\tau}\left(1- \Lambda_{\sigma\tau}G_\sigma G_\tau\right)
  \left[\gamma_{\sigma\tau}  \langle G_\sigma^2\rangle \langle G_\tau^2
    \rangle -\gamma_{\sigma\sigma} \langle G_\sigma^2\rangle^2
    -\gamma_{\tau\tau} \langle G_\tau^2\rangle^2\right]}\ ,
\end{eqnarray}
where $\langle G_{\sigma }G_{\tau }\rangle $ is defined as in
(\ref {eq:mf_cond}) and $\gamma _{\sigma \tau }=\gamma
(E_{F}+i\sigma 0^{+},E_{F}+i\tau 0^{+})$. Here we used the fact
that the velocity can be factorized, and $\langle v_{\alpha
}^{2}\rangle =t^{*2}/2d$ on a hypercubic
lattice in the high-dimensional limit. We immediately see that expression (%
\ref{eq:cond_dos}) is exact to $O(1/d)$ for the conductivity and to $%
O(1/d^{2})$ concerning the vertex corrections. For the conductivity itself
to be exact to $O(1/d^{2})$ an additional contribution in the numerator due
to self-energy corrections of $O(1/d)$ would have to be included. This,
however, would not affect the vertex corrections in the leading order. We
note once more that results obtained from a direct expansion of the
conductivity or the self-energy in powers of $1/d$ would {\em not} be
guaranteed to be physical, e.g., the conductivity may become negative and
the self-energy non-Herglotz.

Vertex corrections to the conductivity are particularly important in low
dimensions ($d=1,2$), where they lead to Anderson localization even for
arbitrarily weak disorder. This is due to the diffusion pole in the
two-particle propagator with energies from different complex half-planes.
Eq.~(\ref{eq:mf_cond}), which was derived in the limit of high dimensions,
is not expected to be applicable in low dimensions, or to describe Anderson
localization. Nevertheless, it is apparent from the form of the vertex in
eq.~(\ref{eq:decoupling}) that $\Lambda ^{eh}$ includes a diffusion pole
(in $\Gamma ^{ee}$) and thereby also a Cooper pole describing coherent back
scattering. For $(-\sigma \tau )=+1$ the ${\bf k}$-integrals over the bare
Cooper poles\ in (\ref{eq:Lambda_prime}) diverge in dimensions $d\leq 2$.
Hence $d=2$ naturally appears as a critical dimension in our high
dimensional approximation where the mean-field description must break down.

\begin{figure}
\noindent
  \epsfig{file=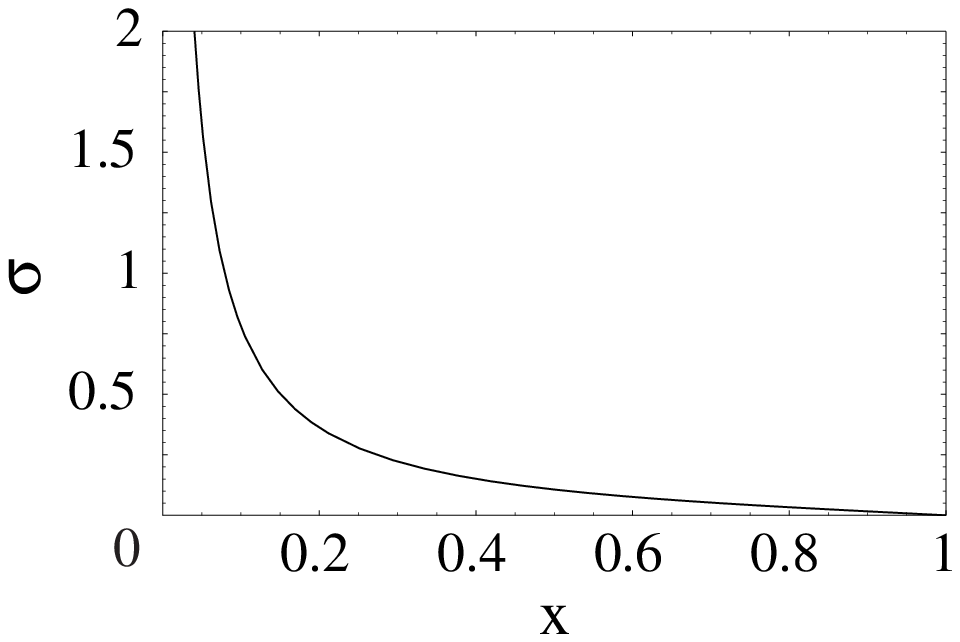, height=4.4cm}
  \epsfig{file=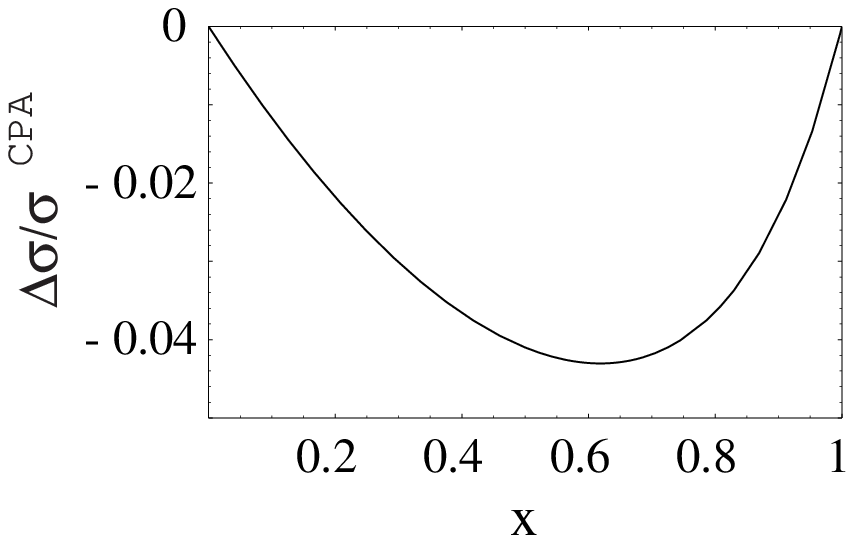,  height=4.4cm}
\caption{Mean-field conductivity with vertex corrections
  $\sigma=\mbox{Re}\ \sigma^{\text{perc}}_{\alpha\alpha}$ from
  (\ref{eq:cond_cheese}) and relative difference
  $\Delta\sigma/\sigma^{\text{CPA}}=\mbox{Re}\ \sigma^{\text{perc}}_{\alpha\alpha}/
  \mbox{Re}\ \sigma^{\text{CPA,perc}}_{\alpha\alpha} -1$ as a function of the
  concentration $x$ for percolation disorder in $d=3$.}
\label{fig:conductivity_perc}
\end{figure}

In dimensions $d>2$ the effect of the diffusion pole in $\Gamma ^{ee}$ on
the electrical conductivity is not as strong in general and depends on the
strength and type of disorder, band filling and the lattice structure. Below
we demonstrate this trend with the help of two different disorder
distributions. We put the Fermi energy into the band center and choose the
dispersion of a $d=\infty $ Bethe lattice, with next-neighbor hopping
amplitude $t^{\ast }=1$ to simplify the relation between the self-energy and
the local propagator. First we choose the simplest disorder model, i.e. a
percolation-type disorder distribution with $V_{i}=0,\infty $ occuring with
probabilities $1-x,x$, respectively. The conductivity from (\ref{eq:cond_dos}%
) can now be evaluated explicitly. The result is
\begin{subequations}\label{eq:cond_cheese_both}
\begin{eqnarray}\label{eq:cond_cheese}
\mbox{Re}\ \sigma _{\alpha \alpha }^{\text{perc}} &=&
\left( \frac{e^{2}}{4\pi d}\right)\left\{ \frac{\frac{1-x}{x}}{1+\frac{3}{2d}%
\frac{x(1-x)}{(2-x)^{2}}}+\frac{\frac{1-x}{2-x}}{1+\frac{1}{2d}\frac{x^{2}(1-x)}
{(2-x)^{3}}}\right\} \ .
\end{eqnarray}
The conductivity (\ref{eq:cond_cheese}) remains positive
everywhere and is a monotonically decreasing function of the
disorder strength $x$. Fig.~\ref{fig:conductivity_perc} shows that
the vertex corrections {\em decrease} the one-particle (i.e. CPA-)
conductivity
\begin{eqnarray} \label{eq:cond_cheese_CPA}
\mbox{Re}\ \sigma _{\alpha \alpha }^{\text{CPA,perc}} &=&
 \left( \frac{e^{2}}{4\pi
d}\right)\frac{2(1-x)}{x(2-x)}.
\end{eqnarray}
\end{subequations}
for all concentration values $x$.

\begin{figure}
\noindent
  \epsfig{file=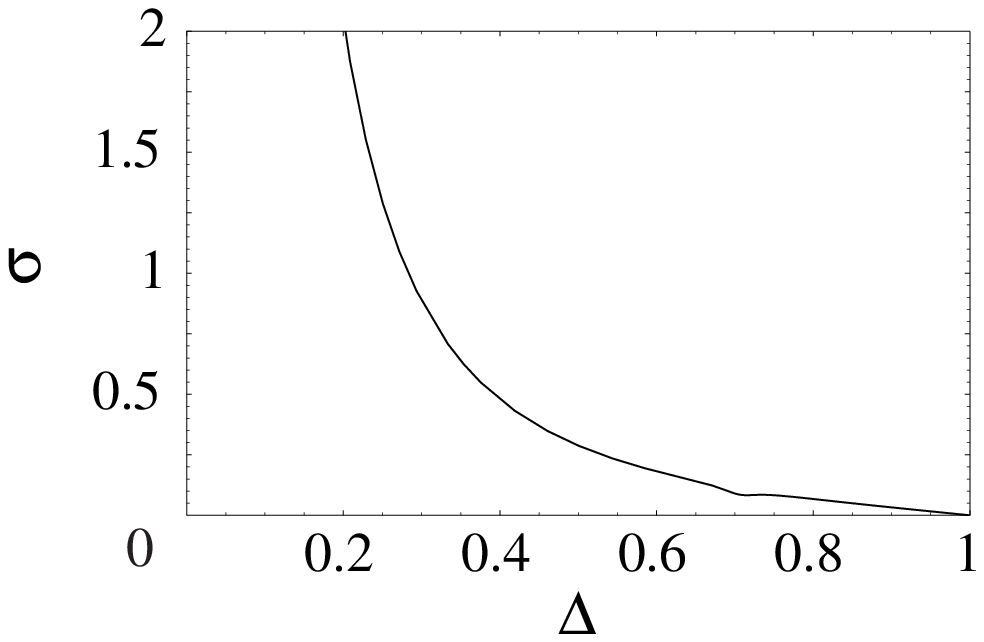, height=4.4cm}
  \epsfig{file=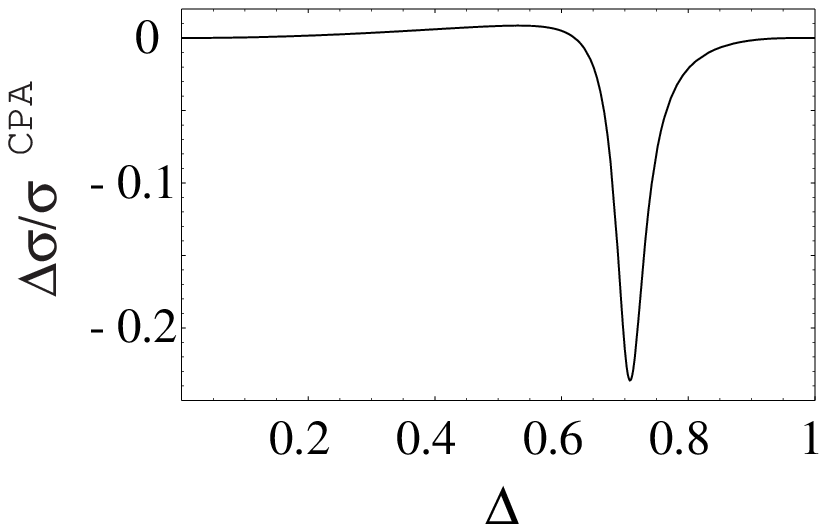,  height=4.4cm}
\caption{Mean-field conductivity with vertex corrections
  $\sigma=\mbox{Re}\ \sigma^{\text{bin}}_{\alpha \alpha}$ from
  (\ref{eq:cond_binary}) and relative difference
  $\Delta\sigma/\sigma^{\text{CPA}}=\mbox{Re}\ \sigma^{\text{bin}}_{\alpha\alpha}/
  \mbox{Re}\ \sigma^{\text{CPA,bin}}_{\alpha\alpha}-1$ as a function of the disorder
  strength $\Delta$ for binary alloy in $d=3$.}
\label{fig:conductivity_bin}
\end{figure}

Another and more important example is the standard binary alloy with two
values of the random potential $V_{i}=\pm \Delta $ with equal probability,
where we obtain
\begin{subequations}\label{eq:cond_binary_both}
\begin{eqnarray} \label{eq:cond_binary}
\mbox{Re}\ \sigma _{\alpha \alpha }^{\text{bin}} &=&
\left( \frac{e^{2}}{4\pi d}\right)\left\{ \frac{\frac{1-\Delta ^{2}}{\Delta ^{2}}%
}{1-\frac{1}{2d}\frac{\Delta ^{2}(1-\Delta ^{2})}{(2-\Delta ^{2})^{2}}}+\frac{%
\frac{1-\Delta ^{2}}{2-\Delta ^{2}}}{1+\frac 1{2d} \frac{\Delta
^{4}(1-\Delta ^{2})}{(1-2\Delta^2)^2(2-\Delta ^{2})^{3}}}\right\} \ .
\end{eqnarray}
Fig.~\ref{fig:conductivity_bin} shows that vertex
corrections cause the conductivity to  slightly {\em increase }
with respect to the CPA result
\begin{eqnarray}\label{eq:cond_binary_CPA}
\mbox{Re}\ \sigma _{\alpha \alpha }^{\text{CPA,bin}} &=&
 \left( \frac{e^{2}}{4\pi d}\right)\frac{2(1-\Delta ^{2})}{\Delta
   ^{2}(2-\Delta ^{2})}.
\end{eqnarray}
\end{subequations}
at weak and moderate disorder strength. Around $\Delta
=1/\sqrt{2}$, the two-particle scattering in the vertex
corrections result in a {\it decrease} of the CPA conductivity.
For stronger disorder conductivity (\ref{eq:cond_binary}) remains
positive and smaller than the CPA result
(\ref{eq:cond_binary_CPA}) up to the split-band limit $\Delta =1$,
where it vanishes. Both conductivities (\ref{eq:cond_binary_both})
monotonically decrease as functions of the disorder strength.
However, the largest relative suppression of the CPA conductivity
due to vertex corrections occurs at $\Delta =1/\sqrt{2}$ where the
denominator of the second term in (\ref{eq:cond_binary}), i.e.,
the terms with $\sigma \tau =1$ in (\ref{eq:cond_dos}), diverges
and changes sign. Note that if the denominators in (\ref
{eq:cond_binary}) were expanded only to the leading order in
$1/d$, the conductivity would turn {\it negative} and hence become
unphysical around $\Delta =1/\sqrt{2}$. This explicitly
demonstrates the importance of calculating the conductivity via
approximations to the irreducible vertex $\Lambda ^{eh}$, as
proposed in this paper, instead of evaluating the conductivity
corrections to the single-bubble diagram directly from the full
vertex $\Gamma $.

The results for the two models presented above demonstrate that
the effect of vertex corrections in high dimensions is non-universal.
Additional scatterings from the random potential contained in the vertex
corrections can decrease as well as increase the single-bubble conductivity.
When treated inappropriately vertex corrections can even lead to
unphysical, negative results. The actual impact of vertex corrections in $d>2
$ depends on band structure, band filling, and disorder distribution and
strength.

In conclusion, we derived a mean-field expression for the vertex corrections
to the electrical {\em dc}-conductivity which becomes exact in the
asymptotic limit of high lattice dimensions. To warrant the conductivity to
be non-negative it was expressed as a functional of the {\em irreducible}
vertex function in the electron-hole channel $\Lambda ^{eh}$. We calculated
the leading high-dimensional asymptotics of the {\em \ non-local} part of $%
\Lambda ^{eh}$ and thereby derived an expression in closed form for the {\em %
dc}-conductivity including vertex corrections. Although this mean-field
approximation does not describe Anderson localization, it goes
systematically beyond the CPA. The result, Eq.~(\ref{eq:mf_cond}), can be
used as a mean-field formula for calculating the effects of vertex
corrections to the electrical conductivity in, e.g., three-dimensional
alloys, and may serve as a starting point for improved approximation schemes
beyond the CPA limit. To include the essentials of the physics of Anderson
localization one needs to improve the mean-field approximation for the
irreducible vertex presented in this paper. A minimal requirement for this
is a self-consistent theory for the non-local part of the vertex $%
\Lambda^{eh}$ which may, for example, be obtained from a parquet
approximation.

This work was supported in part by Grant No. 202/98/1290 of the Grant Agency
of the Czech Republic (VJ) and by the Sonderforschungsbereich 484 of the
Deutsche Forschungsgemeinschaft.


\begin{references}
\bibitem{Velicky68}  B. Velick\'y, S. Kirkpatrick, and H. Ehrenreich, %
\newblock Phys. Rev. {\bf 175}, 745 (1968).

\bibitem{Elliot74}  R. J. Elliot, J. A. Krumhansl, and P. L. Leath, %
\newblock Rev. Mod. Phys. {\bf 46}, 465 (1974).

\bibitem{Mueller-Hartmann73}  E.~M\"uller-Hartmann, \newblock Solid. State
Comm. {\bf 12}, 1269 (1973).

\bibitem{Wegner79}  F. J. Wegner, \newblock Phys. Rev. B{\bf 19}, 783 (1979).

\bibitem{Metzner89}  W. Metzner and D. Vollhardt, \newblock Phys. Rev. Lett.
{\bf 62}, 324 (1989).

\bibitem{Vlaming92}  R. Vlaming and D. Vollhardt, \newblock Phys. Rev. {\bf %
45}, 4637 (1992).

\bibitem{Janis92}  V. Jani\v s and D. Vollhardt, \newblock
  Phys. Rev. B{\bf 46}, 15712 (1992).

\bibitem{Georges96}  A.~Georges, G.~Kotliar, W.~Krauth, and M.~Rozenberg, %
\newblock Rev. Mod. Phys. {\bf 68}, 13 (1996).

\bibitem{Janis99a}  V.~Jani\v s, \newblock Phys. Rev. Lett. {\bf 83}, 2781
(1999).

\bibitem{Hettler99}  M.~H.~Hettler, M.~Mukerjee, M.~Jarrell, and
H.~R.~Krishnamurthy, \newblock Phys. Rev. B{\bf 61}, 12739 (2000).

\bibitem{Khurana90}  A.~Khurana, \newblock Phys. Rev. Lett. {\bf 64}, 1990
(1990).

\bibitem{Note1} Note that the definition of a ''mean-field conductivity''
  is ambiguous, since we can implement the high-dimensional asymptotics in
  two ways. We can either analyze directly the diagrams contributing to the
  conductivity, find out the leading asymptotic contribution in high
  dimensions and define the result as a mean-field conductivity. This will
  lead to the single-bubble diagram of order $1/d$.  Or we may start with
  the two-particle vertex $\Gamma$, take its leading asymptotic form,
  insert it into the Kubo formula for the current-current correlation
  function and identify the resulting expression with a mean-field
  conductivity. The vertex corrections appearing in this mean-field
  conductivity cannot be eliminated by resorting to symmetry.
  \cite{Janis99a} They are of order $O(1/d^{2}) $ and do not change the
  leading high-dimensional asymptotics of the single-bubble diagram. The
  difference between the two definitions of the mean-field conductivity
  becomes important, however, in finite dimensions, where one actually
  wants to use the high-dimensional asymptotics as a mean-field
  approximation.

\bibitem{Lee85}  P.~A.~Lee and R.~V.~Ramakrishnan, \newblock
  Rev. Mod. Phys. {\bf 57}, 287 (1985).

\bibitem{Vollhardt92}  D.~Vollhardt and P.~W\"olfle in {\it Electronic Phase
Transitions}, eds. W. Hanke and Yu. V. Kopaev (Elsevier, Amsterdam 1992),
p.1.

\bibitem{Jarrell+Krish00}  M. Jarrell and H. R. Krishnamurthy, \newblock
  e-print {\it cond-mat}/0006431.

\bibitem{Janis00}  A detailed derivation of the asymptotic vertex functions
for noninteracting electrons in a random potential will be presented in a
separate paper.

\bibitem{Janis99b}  V.~Jani\v s, \newblock Phys. Rev. B{\bf 60}, 11345
(1999).

\bibitem{Velicky69}  B.~Velick\'y, \newblock Phys. Rev. {\bf 184}, 614
(1969).

\bibitem{Mueller-Hartmann89}  E.~M\"uller-Hartmann, \newblock Z. Phys. B -
Condensed Matter {\bf 74}, 507 (1989).
\end{references}
\end{document}